\documentclass[iop]{emulateapj}

\usepackage{amsmath}
\newcommand{\alephuv}{f$_{\mathrm{esc}}$(0.16$\,\micron$)}
\newcommand{\alephha}{f$_{\mathrm{esc}}$(H$\alpha$)}
\newcommand{\alephuvha}{f$_{\mathrm{esc}}$(0.66$\,\micron$)}
\newcommand{\alephhb}{f$_{\mathrm{esc}}$(H$\beta$)}

\begin{document}

\title{Dusty galaxies and the degeneracy between their dust distributions and
the attenuation formula}

\author{Kyle Penner\altaffilmark{1}, Mark Dickinson\altaffilmark{2},
Benjamin Weiner\altaffilmark{3}, Hanae Inami\altaffilmark{2},
Jeyhan Kartaltepe\altaffilmark{2}, Janine Pforr\altaffilmark{4},
Hooshang Nayyeri\altaffilmark{5},
Susan Kassin\altaffilmark{6},
Casey Papovich\altaffilmark{7}, and
Alexandra Pope\altaffilmark{8}
}

\email{kdpenner@gmail.com}

\altaffiltext{1}{Laboratoire AIM, DSM/Irfu/SAp, CEA-Saclay,
91191 Gif-sur-Yvette, France}
\altaffiltext{2}{National Optical Astronomy Observatory,
Tucson, AZ 85719, USA}
\altaffiltext{3}{Department of Astronomy, University of Arizona,
Tucson, AZ 85721, USA}
\altaffiltext{4}{Laboratoire d'Astrophysique de Marseille, Aix
Marseille Universit\'e, 13388 Marseille cedex 13, France}
\altaffiltext{5}{Department of Physics \& Astronomy, University of California
Irvine, Irvine, CA 92697, USA}
\altaffiltext{6}{Space Telescope Science Institute, Baltimore, MD 21218, USA}
\altaffiltext{7}{Department of Physics \& Astronomy, Texas A\&M University,
College Station, TX 77843, USA}
\altaffiltext{8}{Department of Astronomy, University of Massachusetts Amherst,
Amherst, MA 01003, USA}

\begin{abstract}
Do spatial distributions of dust grains in galaxies have typical forms, as do
spatial distributions of stars?  We investigate whether or not the distributions
resemble
uniform foreground screens, as commonly assumed by the high-redshift galaxy
community.  We use rest-frame infrared,
ultraviolet, and H$\alpha$ line luminosities of dust-poor and dusty galaxies
at $z \sim 0$ and $z \sim 1$ to compare measured H$\alpha$ escape fractions
with those predicted by the Calzetti attenuation formula.  The predictions,
based on
UV escape fractions, overestimate the measured H$\alpha$ escape fractions for
all samples.  The
interpretation of this result for dust-poor $z \sim 0$ galaxies is that regions
with ionizing stars have more dust than regions with nonionizing UV-emitting
stars.  Dust distributions for these galaxies are nonuniform.  The
interpretation of the overestimates for dusty galaxies at both redshifts is
less clear.  If the attenuation formula is inapplicable to these galaxies,
perhaps the disagreements are unphysical; perhaps dust distributions in
these galaxies are uniform.  If the attenuation formula does apply, then dusty
galaxies have nonuniform dust distributions; the distributions are more
uniform than they are in dust-poor galaxies.  A broad range of H$\alpha$ escape
fractions at a given UV escape fraction for $z \sim 1$ dusty galaxies, if real,
indicates diverse dust morphologies and the implausibility of the screen
assumption.
\end{abstract}

\section{Introduction}

A galaxy's morphology is often defined as its spatial distribution of stars.
Rest-frame optical images of galaxies show diverse arrangements of
their stellar components: smooth ellipticals, tightly-wound spirals, clumpy
disks, barred disks, train-wrecks, fuzzballs, and whatever Willman 1 is
\citep{willman05,willman11}.
Little more than spatial extent is known about distributions of dust in dusty
galaxies
\citep{younger10,diaz10,simpson15}.  Yet spatial distributions of
all galactic components have intrinsic value.  If we want to learn about
galaxies, morphology should be a wavelength-independent word.  Surprises
surely abound: for example, galaxies at $z \sim 0.3$ have dust out to several
Mpc.  Their average dust mass profile is a constant fraction of their average
halo mass profile \citep[][see also \citealt{zaritsky94}]{menard10}.

Dust attenuates a galaxy's intrinsic luminosity at ultraviolet (UV) and optical
wavelengths.  To recover an intrinsic luminosity from an emergent luminosity we
often assume that dust grains
are distributed as a uniformly thick screen between us and the galaxy's stars
\citep{calzetti94}.  From the uniform screen assumption follows an
attenuation formula that depends solely on wavelength.

An attenuation formula is denoted $k(\lambda)$ or $k'(\lambda)$.  It differs
from an extinction formula, which is independent of the spatial distribution
of dust.  When we derive an extinction formula, it is valid for the small area
surrounding a star in the Milky Way, not the large area containing a mixture of
stars in distant galaxies.

If we write this equation:
\begin{equation*}
L_{\mathrm{emergent}}/L_{\mathrm{intrinsic}} = 10^{-\mathrm{A}/2.5} = 10^{-\mathrm{E(B-V)} k(\lambda)/2.5}
\end{equation*}
we assume a uniform screen not because $k(\lambda)$ is solely dependent on
wavelength, but because $L_{\mathrm{emergent}}/L_{\mathrm{intrinsic}}$ is solely
dependent on
wavelength.  In an illustrative sense
$L_{\mathrm{emergent}}/L_{\mathrm{intrinsic}}$
is an attenuation formula convolved with a spatial distribution of dust.

The uniform screen assumption is unrealistic.  A low-redshift galaxy with a
high emergent UV luminosity and a low IR luminosity---a dust-poor
galaxy---has a high UV escape fraction.  The \citet{calzetti00} formula
predicts its H$\alpha$ escape fraction; the prediction
overestimates the measured H$\alpha$ escape fraction.  The galaxy's regions with
$>10$ M$_{\odot}$ stars, which ionize gas, have more dust than its regions
with less massive stars that emit in the UV and leave the gas unionized
\citep{calzetti94,calzetti97b}.  While the discrepancy in amounts of dust does
not invalidate the screen part of our assumption, it does invalidate the
uniform part.

The screen part of our assumption is suspect, but it may be a
good approximation to the truth---the alternative assumption being that dust
mixes with stars.  \citet{liu13} study the dust distribution of M83 at a
spatial scale of 6 pc.  They divide M83 into a center and outskirts.  For each
part, they calculate the percentage of regions with ionizing stars mixed with
dust.  Only in the center does this percentage exceed 50.  If they degrade the
spatial resolution to 100 pc, they conclude that the dust distribution around
ionizing stars is a screen.  The screen assumption is qualitatively similar to
a scenario in which dust is distributed in a large number of clumps
\citep{calzetti94}.  If we choose to think of dust distributions as clumpy, the
number of clumps surrounding nonionizing UV-emitting stars in low-redshift
dust-poor galaxies is 60\% of the number of clumps surrounding ionizing stars
\citep{calzetti97a}.

The discrepancy between the uniform assumption and real dust
distributions in low-redshift dust-poor galaxies grows even larger if we
consider that galaxies have centers and outskirts---that galaxies are not
point sources.  A number of studies find
that H$\alpha$ and UV escape fractions increase with increasing radius
\citep{boissier04,prescott07,munoz09}.  Ionizing stars are surrounded by more
dust than nonionizing UV-emitting stars; ionizing stars
at large radii have less dust than ionizing stars at small radii, and similarly
for nonionizing UV-emitting stars.

We know little about dust distributions in low-redshift dusty galaxies, which
have high IR luminosities and low-to-high emergent UV luminosities.  Their
IR emitting regions span a range of sizes, from sub- to several kpc
\citep{diaz10}.  Measuring a size is hard enough; imaging a distribution is
even harder.  If a dusty galaxy's measured H$\alpha$ escape fraction agreed with
the prediction from the \citet{calzetti00} attenuation formula, we might
conclude that the galaxy has a uniform dust distribution only if we believe
that the prediction is valid.  If a dusty galaxy's measured H$\alpha$
escape fraction disagreed with the prediction from the \citet{calzetti00}
formula, the galaxy might have a nonuniform dust distribution, like that of a
dust-poor galaxy.  This explanation for the disagreement is not unique.  The
galaxy might have a uniform distribution and obey a different $k(\lambda)$.

We have nothing more than vague and conflicting ideas of dust distributions in
all high-redshift galaxies.  \citet{onodera10}, \citet{kashino13}, and 
\citet{price13} argue for nonuniform distributions in their samples, which
comprise dusty and dust-poor galaxies; \citet{erb06} present evidence for
uniform distributions in a similarly mixed sample.  \citet{reddy15} argue
that the uniformity of the dust distribution depends on the galaxy's
star formation rate: as the star formation rate increases the dust distribution
becomes less uniform.

In this paper, we ask the question, For dusty galaxies at $z \sim 0$ and
$z \sim 1$, how do H$\alpha$ escape fractions relate to the predictions made by
the \citet{calzetti00} attenuation formula?  We: (1) show that H$\alpha$
escape fractions differ from
the prediction; (2) show that the relations between H$\alpha$ and UV
escape fractions differ from the relation for low-redshift dust-poor galaxies; and
(3) argue that an interpretation in the context of dust distributions relies on
the shaky assumption that the \citet{calzetti00} attenuation formula is
universally valid.

For this paper we assume a cosmology with $H_{0} = 70$ km s$^{-1}$ Mpc $^{-1}$,
$\Omega_{\mathrm{m}} = 0.3$, and $\Omega_{\Lambda} = 0.7$.  Our terminology
is: (1) a \emph{uniform} screen is a screen equally thick between a region with
stars which ionize gas and a region with stars which emit in the UV and do not
ionize gas; (2) a \emph{dust-poor} galaxy generally has high values of UV and
H$\alpha$ escape fractions; and (3) a \emph{dusty} galaxy is selected from an IR
image and generally has low values of UV and H$\alpha$ escape fractions.  The
samples of dust-poor and dusty galaxies are not
dichotomous.  More details are in the following section.

\section{Data}\label{sec:data}

Measured and derived quantities are in Tables \ref{tbl:highz} and
\ref{tbl:lowz}.

\subsection{Measured quantities}

We use IR, emergent UV, and H$\alpha$ luminosities to determine escape
fractions.  In this section we detail our catalogs and samples.

\subsubsection{Sample of dusty galaxies at $z > 0.7$}

Our study uses observations of the GOODS-S, COSMOS, and UDS regions.  A
catalog of \emph{Herschel}/PACS 100$\,\micron$ sources for GOODS-S comes from
\citet{magnelli13}; for COSMOS and UDS, we use catalogs produced for the
\emph{Herschel} survey of CANDELS regions (Inami et al., in prep).  A
100$\,\micron$ source is defined as a $\ge 3\sigma$ flux density measurement
from PSF fitting to a \emph{Spitzer}/MIPS 24$\,\micron$ source, which is based
on an IRAC 3.6$\,\micron$ source prior.  In GOODS-S, the catalog has
flux densities at 24, 70, 100, and 160$\,\micron$; in COSMOS and
UDS, the catalogs have flux densities at 24, 100, 160, 250, 350,
and 500$\,\micron$.  All flux densities are $\ge 3\sigma$.  We do not use the
250, 350, and 500$\,\micron$ flux
densities in the COSMOS and UDS catalogs, to avoid the effects of source
confusion on PSF fitting.  Including them changes our results negligibly.

We associate each 100$\,\micron$ source with a 1.6$\,\micron$ ($H$-band) source
in the
catalogs produced by CANDELS
\citep[Nayyeri et al., in prep;][]{galametz13,guo13}.  Each catalog has
broadband flux densities at many wavelengths, not all of which are common to the
other two.  We require the 1.6$\,\micron$ source to be a unique match
within 0.7$\arcsec$ of the position of the 100$\,\micron$ source's
3.6$\,\micron$ prior.  The match radius is an estimate for maximizing the
number of unique matches while minimizing the number of multiple matches.

The final quantity of interest for each matched source is an H$\alpha$ line
flux or flux limit.  These come from our reduction of observations from the
\emph{HST}/WFC3-IR grism survey 3D-\emph{HST} \citep{brammer12}.
We match each 100$\,\micron$ source to a counterpart in the direct image
accompanying the grism observations.  Two people in our group visually inspect
the counterpart's spectrum and then review and reconcile any discrepant redshift
assignments.  The
H$\alpha$ line falls in the grism's wavelength range for galaxies at
$0.7 < z < 1.5$.  The spectra have low enough wavelength resolution that the
H$\alpha$ and [N II] lines are blended; we assume that the line flux of [NII]
at $\lambda = 0.6583\,\micron$ is 0.3 times the H$\alpha$ line flux and that
the line flux of [N II] at $\lambda = 0.6548\,\micron$ is 0.1 times the
H$\alpha$ line flux.  The ratio of [N II] to H$\alpha$ line fluxes depends on
gas-phase metallicity so these values may be incorrect for some star-forming
galaxies.
The values will be correct for a significant fraction of galaxies in our
sample if the sample has a distribution of
gas-phase metallicity similar to that of low-redshift galaxies
\citep{kauffmann03}.  The values will be correct for a larger fraction of
galaxies in our sample if the sample has a distribution of gas-phase
metallicity similar to that of low-redshift galaxies which are luminous in
the near-IR \citep{weiner07}.  For each 1D spectrum with a visible and
uncontaminated
line complex, we fit a linear model to the continuum and Gaussian profiles to
the emission lines and extract the H$\alpha$ line flux.  The uncertainty on the
line flux comes from the covariance matrix times the fit's reduced $\chi^{2}$.
We replace $< 5\sigma$ H$\alpha$ line fluxes with limits.  We multiply the line
fluxes and limits by 1.1 as an aperture correction.

Some sources will be at $0.7 < z < 1.5$ but will not have visible H$\alpha$
line flux.  To determine line flux upper limits for these sources, we match each
100$\,\micron$ source to a source with a spectroscopic redshift in
the catalogs of: \citet{lefevre04,szokoly04,mignoli05,yamada05,ravikumar07,
vanzella08,lilly09,balestra10,fadda10,simpson12,kurk13,santini14}; Simpson et
al. (in prep.); Almaini et al. (in prep.); and
M. Dickinson (private communication).  Operationally both the
100$\,\micron$ and spectroscopic catalogs are matched to the CANDELS catalogs.
If the 100$\,\micron$ source has a spectroscopic redshift that is unconfirmed
by an uncontaminated grism spectrum, we estimate the widths of the invisible
Gaussian line profiles from the ratio of integrated flux density to peak
surface brightness for the region of the direct image surrounding the galaxy.
The direct
image is the image taken as part of the grism observations.  We then fit
the same model as above to the 1D spectrum and extract the H$\alpha$ line flux
uncertainty.  We use $5\sigma$ limits.  We are unable to determine limits for
galaxies observed by the grism that are undetected in the direct image.  Most
limits are in GOODS-S because the spectroscopic completeness is higher there
than it is in COSMOS or UDS.

To summarize, our high-redshift sample contains galaxies with: (1) detected
emission at 100$\,\micron$; (2) detected emission at 1.6$\,\micron$; and (3)
spectroscopically-determined redshifts at $0.7 < z < 1.5$.

\subsubsection{Samples of dust-poor and dusty galaxies at $z < 0.2$}

We use a nonstandard method to determine H$\alpha$ escape fractions, so we test this
method first for low-redshift dust-poor galaxies which have known dust
properties.  We choose a sample of star-forming galaxies at $z \sim 0$ that are
analogs of common star-forming galaxies at high redshift
\citep[LBAs;][]{overzier09,overzier11}.

We use 24$\,\micron$ flux densities in the \emph{Spitzer} Heritage Archive.  Of
the 26 sources in \citet{overzier11}, 3 have unreliable 24$\,\micron$ flux
densities per the recommendations in the enhanced imaging products quick start
guide; these sources are excluded.  The remaining 23 galaxies have flux
densities at
0.15 and 0.23$\,\micron$ (NUV and FUV bands), from \emph{GALEX} observations.
H$\alpha$ line
fluxes, from SDSS observations, are from Overzier (private communication).  We
multiply the line fluxes by 1.67 as a fiber-aperture correction
\citep{overzier09}.

We also apply our method to a sample of low-redshift dusty galaxies, which have
relatively unknown dust properties.  \citet{hwang13} collate, for low-redshift
dust-obscured galaxies (DOGs) and a control sample: flux densities at
9, 12, 22, 25, 60, 100, and 140$\,\micron$, from \emph{IRAS}, \emph{AKARI}, and
\emph{WISE} observations; 0.15 and 0.23$\,\micron$ flux densities, from
\emph{GALEX} observations; and H$\alpha$ line fluxes, from SDSS observations.
The fiber-aperture corrections for the H$\alpha$ line fluxes come from the
differences, in the accompanying 0.62$\,\micron$ ($r$-band) images, between
Petrosian flux
densities and flux densities in apertures with equal size to the fiber
apertures.

\subsection{Derived quantities}

In this section we detail how we estimate total IR and emergent UV
luminosities, UV continuum power-law indices, and star formation rates for the
galaxies in our samples.

\subsubsection{IR luminosities}

We estimate a total IR luminosity (8--1000$\,\micron$; $L_{\mathrm{IR}}$) for
each galaxy.  We redshift the \citet{chary01} template spectral energy
distributions (SEDs) to the distance of each galaxy.  If the galaxy has one
measured IR flux density, we find the SED that most closely matches that flux
density and multiply the IR luminosity of the SED by the ratio between actual
and predicted flux densities.  If the galaxy has two or more measured IR flux
densities, we introduce a multiplicative factor $f$ for each SED.  We find the
best-fit $f$ for each SED; the result is an $f$ and $\chi^{2}$ value.  We
choose the SED and its accompanying $f$ that results in the
minimum of all $\chi^{2}$ values.  The galaxy's $L_{\mathrm{IR}}$ is then $f$
times the IR luminosity of the SED.  In practice, all the high- and
low-redshift dusty galaxies have two or more measured IR flux densities, and
all the low-redshift dust-poor galaxies have one measured IR flux density.

\citet{overzier11}, in their footnote 14, find that the \citet{chary01} SEDs do
not fit low-redshift dust-poor galaxies as well as other SEDs.  They measured
IR flux density limits at 70 and 160$\,\micron$ as well as flux densities at
24$\,\micron$.  The footnote explains that the \citet{chary01} SEDs, when
fit to 24$\,\micron$ flux densities, predict flux densities at 70 and
160$\,\micron$ that fall above their upper limits.  Our IR luminosities are
high compared to theirs---on average, a factor of 1.4 times higher; at most, a
factor of 2.5 times higher.  We present results for both sets of IR
luminosities for the low-redshift dust-poor galaxies.

IR luminosities are between $5\times10^{10}$ and $1\times10^{12} L_{\odot}$
for the central 95\% of the sample of high-redshift dusty galaxies; between
$7\times10^{10}$ and $4\times10^{11} L_{\odot}$ for the low-redshift dusty
galaxies; and between $4\times10^{10}$ and $5\times10^{11} L_{\odot}$ for
the low-redshift dust-poor galaxies.

\subsubsection{UV continuum power-law indices and luminosities}

For galaxies with UV emission from newly formed massive stars the continuum
between 0.125 and 0.263$\,\micron$ is approximately a power law with index
$\beta$:
\begin{equation}
S_{\lambda} = C\lambda^{\beta}
\end{equation}
where $S_{\lambda}$ is flux density and $\lambda$ is wavelength.
We use all measured flux densities and uncertainties in this rest-frame
wavelength range to fit for $\beta$ and the constant factor $C$ or
for their limits.  Measured flux densities may be negative---the sky aperture
may be brighter than the galaxy aperture---which
complicates our fitting procedure.  When a galaxy has only two measured flux
densities in the wavelength range, the solution for $\beta$ and $C$ is
analytic, so for a solution to exist the flux densities must be positive.  If
they both are, we make no cuts on their signal-to-noise ratio (S/N).
If the only positive flux density, among multiple measurements, is $> 3\sigma$,
we calculate a $\beta$ limit using that flux density and a $3\sigma$ limit for
another in the rest-frame wavelength range.  Whether the limit is an upper or
lower one depends on whether the detected flux density's wavelength is lower
or higher than the limit's wavelength.  We do not estimate a $\beta$ limit when
the only positive flux density, among multiple measurements, is $< 3\sigma$.
If more than two measured flux densities are positive, regardless of their S/N
and whether the surplus measurements are negative or positive, we minimize
$\chi^{2}$.

We estimate an emergent UV luminosity or a
limit at rest-frame 0.16$\,\micron$ for each galaxy using its redshift and
the power-law fit to the rest-frame UV flux densities.  We refer to the
luminosity as $0.16 \times L_{0.16}$.

\subsubsection{Star formation rates}\label{sfrs}

The equations in \citet{kennicutt98} relate IR, emergent UV, and emergent
H$\alpha$ luminosities to star formation rates (SFRs).  We assume
that a galaxy's total SFR: (1) is the sum of its IR- and emergent UV-derived
SFR; and (2) equals the intrinsic (not emergent)
H$\alpha$-derived SFR.  The SFR equations are valid under different assumptions
for the duration of a galaxy's star formation episode.  Our assumptions will not
conflict with those of \citet{kennicutt98} if: (1) the SFR has been
constant for $\sim 10^{8}$ yr, so that the instantaneous H$\alpha$-derived SFR
equals the prolonged UV- and IR-derived SFRs; and (2) the dominant
source for the IR luminosity is not old stars or an active nucleus.

We must exclude galaxies with active nuclei because fractions of their IR,
emergent UV, and emergent H$\alpha$ luminosities will be unrelated to newly
formed massive stars.  We identify 5 AGN hosts among the high-redshift dusty
galaxies; we use the \citet{donley12} criteria, which are based on mid-IR flux
densities, which all 3 CANDELS catalogs have.  These criteria will miss some
AGN hosts \citep{juneau13}.  We accept the contaminants because in some fields
we lack data---for example, X-ray data---that we could use to identify the
missed AGN hosts.

Among the low-redshift dusty
galaxies, we keep only those identified as star-forming by \citet{hwang13}.
They use rest-frame optical line flux ratios to discriminate between an AGN host
and a star-forming galaxy without an AGN.  None of the low-redshift dust-poor
galaxies host an AGN \citep{overzier11}.

\section{Results}\label{sec:results}

Three star formation rates allow us to compare two attenuation values.
We eschew magnitude attenuation in favor of the escape fraction, which
is the ratio of emergent to intrinsic luminosity.  We define the following
quantities.
\begin{itemize}
\item \alephuv~is the escape fraction at rest-frame 0.16$\,\micron$.  In
relation to A$_{\mathrm{UV}}$,
f$_{\mathrm{esc}}$(0.16$\,\micron$) = 10$^{-\mathrm{A}_{\mathrm{UV}}/2.5}$.
\item \alephha~is the escape fraction of the H$\alpha$ line luminosity.
\item \alephuvha~is the escape fraction extrapolated from \alephuv~to the
wavelength of H$\alpha$---the attenuation formula prediction.  For the
attenuation formula in \citet{calzetti00}, \alephuvha~= \alephuv$^{0.33}$.
\end{itemize}

\citet{calzetti94} argue that, in low-redshift dust-poor galaxies, regions with
ionizing stars are spatially separate from regions with nonionizing
UV-emitting stars.  In this framework \alephuvha~is the escape fraction
of the 0.66$\,\micron$ continuum emission from low-to-high mass stars and
\alephha~is the escape fraction of the 0.66$\,\micron$ line emission from gas
ionized by high mass stars.  These galaxies have \alephha~= \alephuv$^{0.69}$,
equivalent to \alephha~= \alephuvha$^{2.1}$ \citep{calzetti94,calzetti97b}.
Because the ionizing and nonionizing stars are spatially separate, the
escape fractions disagree because the dust distributions are nonuniform.  The
ionizing stars are surrounded by more dust than are the nonionizing stars.

The H$\alpha$ escape fractions in \citet{calzetti94} come from the ratios of
H$\alpha$ to H$\beta$
line luminosities.  In the absence of dust the ratio is 2.86, so
($L_{\mathrm{H}\alpha}$/$L_{\mathrm{H}\beta}$)/2.86~=~\alephha/\alephhb.  Since
\alephhb~is a power-law function of \alephha, the ratio of line luminosities
uniquely determines \alephha.

The UV and 0.66$\,\micron$ escape fractions in \citet{calzetti94} come from
comparing spectra of
galaxies with H$\alpha$ escape fractions of 1 to spectra of galaxies with
H$\alpha$ escape fractions less than 1.  The H$\alpha$ and H$\beta$ lines in
the spectra must be masked or replaced with continuum values.  The literature
relation is technically
\alephha~= \alephuvha$^{2.3}$.  \citet{calzetti97b} uses the Milky Way
extinction formula of \citet{howarth83} to determine H$\alpha$ escape
fractions for low-redshift dust-poor galaxies; if we instead use the
attenuation formula, we get \alephha~= \alephuvha$^{2.1}$.

Our hypothesis is that the H$\alpha$ and 0.66$\,\micron$ escape fractions
agree---that dusty galaxies have uniform dust distributions.
Our method for determining H$\alpha$ escape fractions is nonstandard; we lack
H$\beta$ luminosities for the $z \sim 1$ dusty galaxies.  We first apply
it to a sample of low-redshift dust-poor galaxies with the aim of reproducing
the \citet{calzetti97b} result.

\subsection{Reproducing the relation between escape fractions for dust-poor
galaxies}

The sample, from \citet{overzier11}, contains galaxies that lie roughly on
the low-redshift IRX-$\beta$ relation.  This relation is a blunt instrument we
use to determine whether or not galaxies obey an attenuation formula at UV
wavelengths; \alephuv~is a function of IRX and an attenuation formula relates
\alephuv~to $\beta$.  Since these galaxies lie on the relation, we
expect them to have dust properties that conform with the properties of
the galaxies in the \citet{calzetti00} sample.  We calculate \alephha~and
\alephuv~for each galaxy in the following way.
\begin{itemize}
\item \alephuv~= ($L_{\mathrm{IR}}$/($1.68\times0.16\times
L_{\mathrm{0.16}}$)+1)$^{-1}$.  This is equivalent to \alephuv~= (emergent UV
SFR)/(IR SFR + emergent UV SFR).
\item Following the assumptions in \ref{sfrs}: \alephha~=
(emergent H$\alpha$ SFR)/(IR SFR + emergent UV SFR).
\end{itemize}
We use these quantities to find, for the sample, a best-fit power-law index $q$
for the form \alephha~= \alephuv$^{q}$.

The top left panel of Fig. \ref{fha_fuv} shows that \alephha~= \alephuv$^{0.80}$
for the low-redshift
dust-poor galaxies.  If we use the IR and UV luminosities from
\citealt{overzier11}, \alephha~= \alephuv$^{0.84}$.  We lack
uncertainties on any SFRs so we also perform a robust linear regression,
which requires a fit to the form $\log\,$\alephha~= $q\,\log\,$\alephuv.  It
returns a similar result, as does a bootstrap resampling.  We recover a relation
close to the
\citet{calzetti97b} relation for similar galaxies, so we have some confidence
that our procedure is grossly correct.  We stress, again, that there are
no evolutionary connections between the low- and high-redshift galaxy samples;
we aim only to reproduce the \citet{calzetti97b} result with our nonstandard
method of determining \alephha.

\subsection{A new relation for dusty galaxies at low redshift?}

The top right panel of Fig. \ref{fha_fuv} shows that \alephha~=
\alephuv$^{0.55}$ for the low-redshift
dusty galaxies.  We use uncertainties only on the emergent H$\alpha$-derived
SFR---we treat it as the dependent variable---in the weighted fit.  The relation
for dust-poor galaxies is seemingly inapplicable to dusty galaxies at
low redshift.  In \S\ref{subsec:select} and \ref{sec:discuss} we explain why we
preface `inapplicable' with `seemingly.'

\subsection{A corresponding relation for dusty galaxies at high
redshift?}\label{subsec:highz}

The bottom left panel of Fig. \ref{fha_fuv} shows that \alephha~=
\alephuv$^{0.55}$ for high-redshift
dusty galaxies.  We leave limits out of this fit.  The uncertainty in
the index is very small; this is not surprising given the number of galaxies.
The dispersion about the best-fit relation is large; the central 50\% of
galaxies with H$\alpha$-detected SFRs lies between \alephha~= \alephuv$^{0.45}$
and \alephha~= \alephuv$^{0.66}$.  A survival regression, which incorporates
H$\alpha$ SFR limits in a fit to the form $\log\,$\alephha~=
$q\,\log\,$\alephuv, finds $q = 0.47$, as does a simple linear regression using
only the H$\alpha$-detected SFRs.  None of the galaxies have limits for both
the UV and H$\alpha$ SFRs.  The main effect of including limits in the
regression is to increase the standard error of the sample's best-fit index; a
marginally different index is a secondary effect.  Table \ref{regress}
summarizes the relevant quantities for the main fit to each sample of galaxies.

\begin{deluxetable*}{lllll}
\tablecolumns{5}
\tablewidth{272.42545pt}
\tablecaption{Summary of regressions\label{regress}}
\tablehead{
\colhead{Sample} &
\colhead{$q$} &
\colhead{Std. error of $q$} &
\colhead{Residual std. error} &
\colhead{DOF}
}
\startdata
Low-z dust-poor & 0.799 & 0.094 & 6.0  & 13 \\
Low-z dusty     & 0.548 & 0.029 & 67.0 & 54 \\
High-z dusty    & 0.549 & 0.008 & 6.6  & 320
\enddata
\end{deluxetable*}

\subsection{Issues with the regressions}\label{subsec:select}

\begin{figure*}[!ht]
\centering
\includegraphics[scale=0.4]{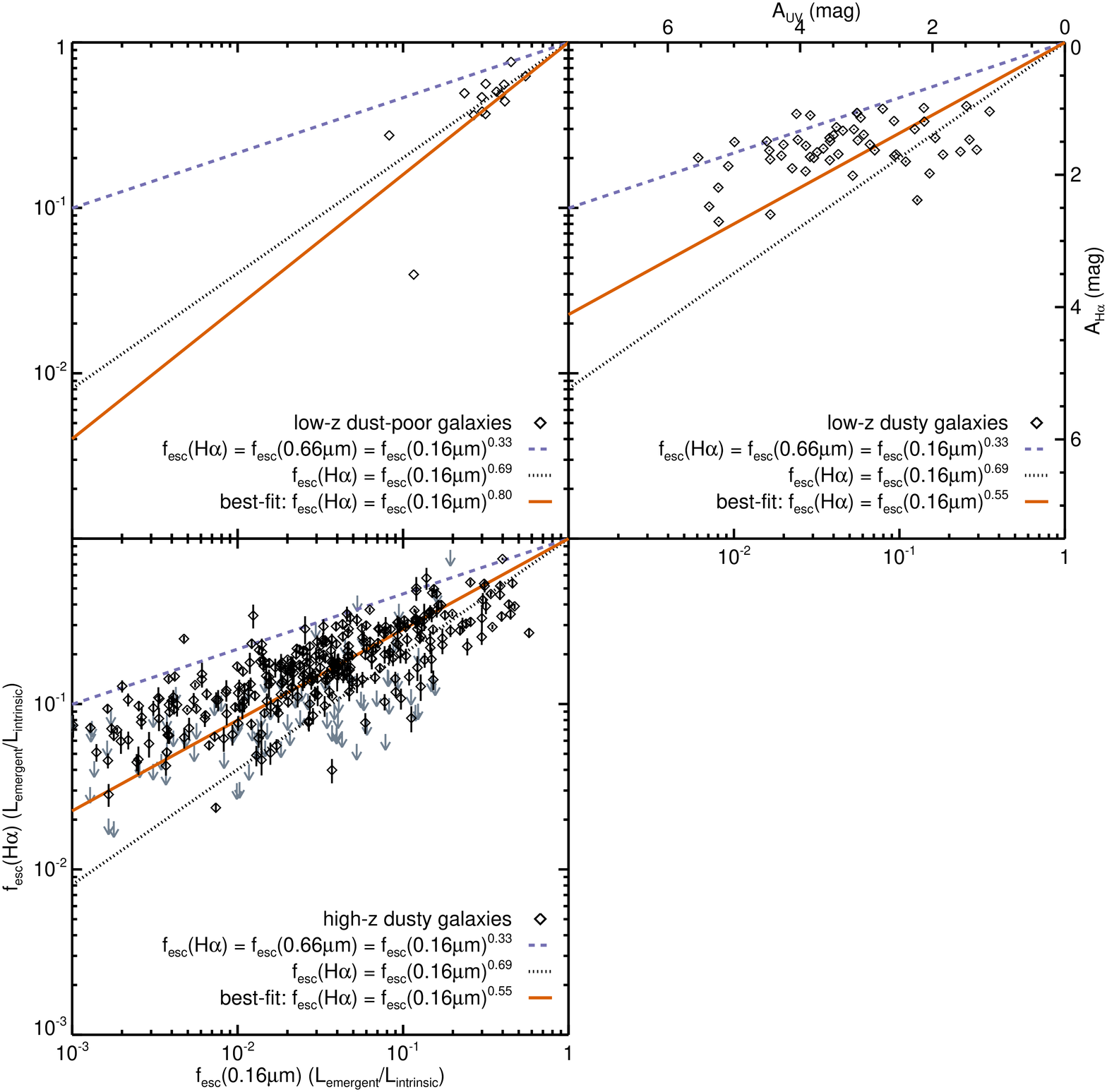}
\caption{{\small \emph{Top left}: The H$\alpha$ escape fraction, \alephha, as a
function of the UV
escape fraction, \alephuv, for the low-redshift
dust-poor galaxies.  The dashed purple line shows \alephuvha, the extrapolation
of \alephuv~to the wavelength of H$\alpha$ using the \citet{calzetti00}
attenuation formula, as a function of \alephuv.  This is the uniform dust
distribution hypothesis for this attenuation formula.  \citet{calzetti97b}, who
study another sample of low-redshift dust-poor galaxies, find
that H$\alpha$ escape fractions are lower than this extrapolation (dotted black
line).  We find a similar result (solid orange line).
We have some confidence that our H$\alpha$ escape fractions are grossly
correct.
\emph{Top right}:  \alephha~as a function of \alephuv~for the low-redshift dusty
galaxies.  In
low-redshift dusty galaxies, regions with ionizing stars have lower
escape fractions than the extrapolation and higher escape fractions than they
would have in low-redshift dust-poor galaxies.
\emph{Bottom left}:  \alephha~as a function of \alephuv~for the high-redshift
dusty galaxies.  In high-redshift dusty galaxies,
regions with ionizing stars have
lower escape fractions than the extrapolation and higher escape fractions than
they would have in low-redshift dust-poor galaxies.  A conclusion about the
uniformity of dust distributions in high-redshift dusty galaxies is subject to
an assumption about the attenuation formula.\label{fha_fuv}}}
\end{figure*}

The escape fraction relations of dusty galaxies are subject to selection
effects.  The best-fit relation for the low-redshift dusty galaxies seems to:
(1) underestimate H$\alpha$ escape fractions at low UV escape fractions; and (2)
overestimate H$\alpha$ escape fractions at high UV escape fractions (top
right panel of Fig. \ref{fha_fuv}).  This may reflect our constraint that
\alephha~=1 when
\alephuv~=1, but the lack of galaxies with H$\alpha$ escape fractions at or
below the relation may be due to galaxies with H$\alpha$ SFRs below the limit
of the sample.  We do not measure H$\alpha$ line flux limits for galaxies that
meet the selection criteria of \citet{hwang13} and have undetected H$\alpha$
line fluxes.  If we replace all H$\alpha$-detected SFRs with SFRs corresponding
to the minimum H$\alpha$ line flux of the sample, we fill in the space around
the relation.

High-redshift dusty galaxies with UV escape fractions $< 10^{-2}$ seem to have
less dispersion in their H$\alpha$ escape fractions than do galaxies with UV
escape fractions $> 10^{-2}$.  Despite our best efforts to measure H$\alpha$
line flux limits, we still lack them for galaxies lacking spectroscopic
redshifts.  If we replace all H$\alpha$-detected SFRs with SFRs corresponding
to a reasonable H$\alpha$ line flux limit for the grism observations
($5 \times 10^{-17}$ erg s$^{-1}$ cm$^{-2}$), we fill in the space with low
H$\alpha$ escape fractions at low UV escape fractions.  The decrease in
dispersion is probably an artifact of the grism detection limit and a paucity
of spectroscopic redshifts in UDS and the part of COSMOS in CANDELS.

\section{Discussion}\label{sec:discuss}

If we assume that dusty galaxies: (1) obey the \citet{calzetti00} attenuation
formula; and (2) were to all have the same UV spectrum in the absence of dust;
our results indicate that their H$\alpha$ and 0.66$\,\micron$ escape
fractions disagree.  Unlike low-redshift dust-poor galaxies, dusty galaxies
have \alephha~= \alephuvha$^{1.6}$; like low-redshift dust-poor galaxies,
dusty galaxies have nonuniform dust distributions.  The relations between
\alephha~and \alephuvha~are similar to those found by \citet{onodera10},
\citet{kashino13}, \citet{price13}, and \citet{reddy15} and conflict with those of
\citet{erb06}.  These studies assume the applicability of the \citet{calzetti00}
attenuation
formula, and use samples of high-redshift galaxies that are not necessarily
dusty.  Their UV escape fractions come from stellar population modeling, while
ours come from IRX; their H$\alpha$ escape fractions come from ratios of
averaged H$\alpha$ to averaged H$\beta$ luminosities, or ratios of H$\alpha$ to
UV luminosities, while ours come from assumptions about SFRs.

However, dusty galaxies may disobey the \citet{calzetti00} attenuation
formula, which complicates our interpretation of the more basic relation
between UV and 0.66$\,\micron$ escape fractions---the relation between
escape fractions for just the regions with nonionizing UV-emitting stars
\citep{reddy10,reddy12,reddy15,buat11,buat12,penner12,kriek13}.  The true attenuation
formula for dusty galaxies may have a higher 0.66$\,\micron$ escape fraction
for a given UV escape fraction than does the \citet{calzetti00} formula.  A
formula modified according to
\citet{kriek13} with $\delta = -0.2$ and no bump at 0.2175$\,\micron$ has
\alephuvha~=
\alephuv$^{0.25}$ instead of \alephuvha~= \alephuv$^{0.33}$.  The combination of
this \citet{kriek13} formula and our result implies that
\alephha~= \alephuvha$^{2.2}$.  The \citep{reddy15} attenuation formula has
\alephuvha~= \alephuv$^{0.23}$, implying that \alephha~= \alephuvha$^{2.4}$.
If the true formula instead has a
much lower 0.66$\,\micron$ escape fraction for a given UV escape
fraction than does the \citet{calzetti00} formula---say, \alephuvha~=
\alephuv$^{0.55}$---we conclude that for dusty
galaxies the H$\alpha$ and 0.66$\,\micron$ escape fractions agree.  Our
conclusion regarding the uniformity of dust distributions depends on the
assumed attenuation formula.  We cannot reject our hypothesis.  The
\citet{calzetti00} formula does demarcate the upper boundary of \alephha~for the
high-redshift dusty galaxies (bottom left panel of Fig. \ref{fha_fuv}), so
if it is the true formula then at least some dusty galaxies have uniform
dust distributions.  The unambiguous statement we make is that dusty galaxies
have relations between H$\alpha$ and UV escape fractions that are different from
the relation for low-redshift dust-poor galaxies.

The dispersion about the best-fit relation for high-redshift dusty galaxies
is large, yet it is susceptible
to our possibly invalid assumption that a galaxy's instantaneous SFR has,
for at least $10^{8}$ yr, equaled its prolonged SFR.  For example, if most
massive stars formed over 15 Myr instead of 300 Myr, we should multiply
$L_{0.16}$ and $L_{\mathrm{IR}}$ by larger numbers than we do here to derive the
UV and IR SFRs.  The true UV SFR is 57\% higher and the true IR SFR is 63\%
higher \citep{madau14}.  The true UV escape fraction is little different;
because the true total SFR is higher, the true H$\alpha$ escape fraction is
lower.  Another scenario is that massive stars stopped forming within the last
15 Myr.  Perhaps some of the high-redshift dusty galaxies have
complicated star formation histories; the combined effect of deriving correct
SFRs might lead to lower dispersion.

\begin{figure}[!ht]
\centering
\includegraphics[scale=0.45]{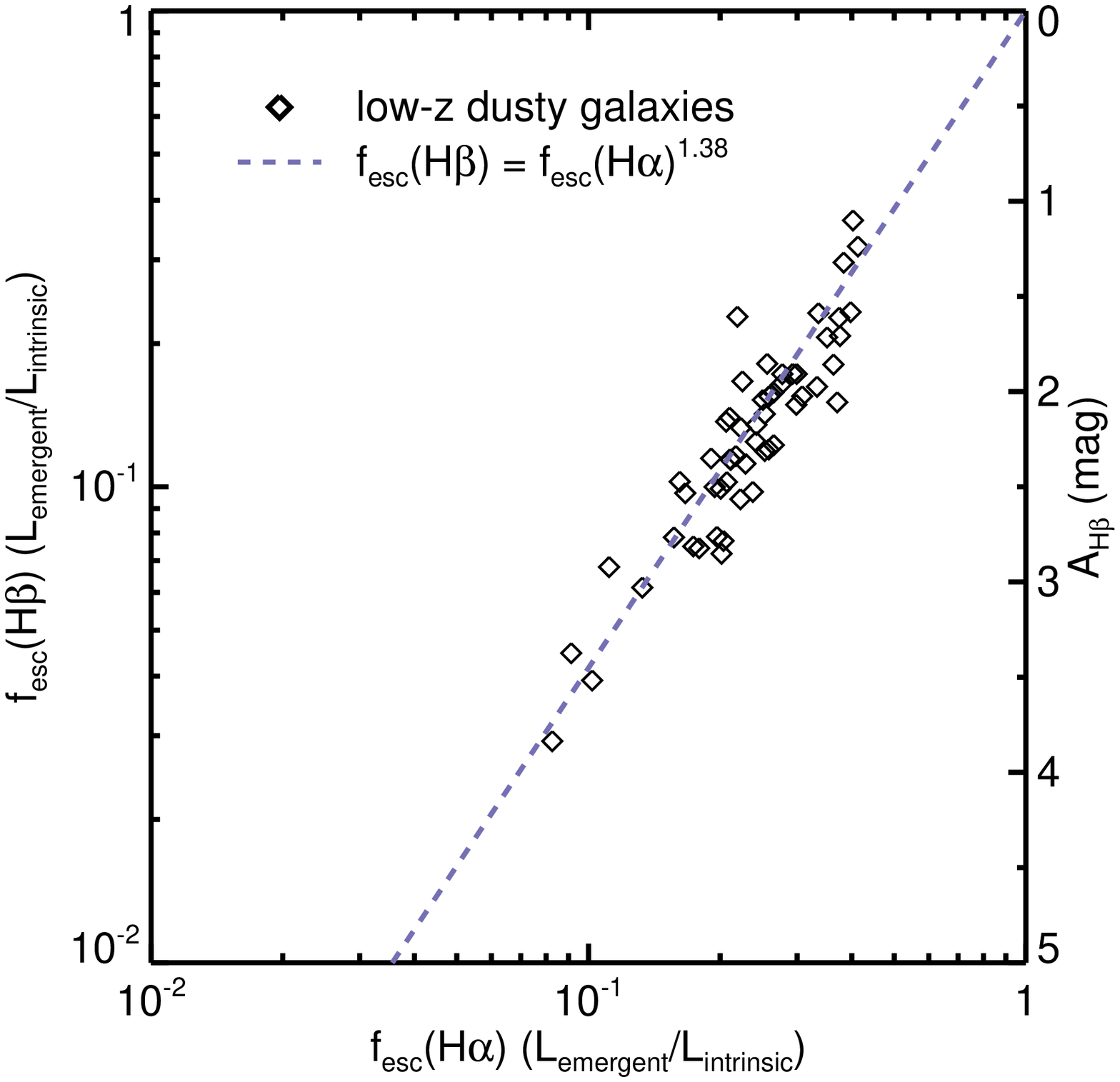}
\caption{H$\beta$ escape fraction, \alephhb, as a function of H$\alpha$ escape
fraction, \alephha, for the low-redshift dusty galaxies.  We determine
\alephhb~using \alephha~and the ratios of H$\alpha$ and H$\beta$ luminosities.
The \citet{calzetti00} attenuation formula has \alephhb~= \alephha$^{1.38}$
(dashed purple line).  If H$\beta$ measurements of high-redshift dusty galaxies
agree with the prediction, we can resolve uncertainties related to star
formation histories and the validity of the \citet{calzetti00} attenuation
formula at optical
wavelengths; otherwise, we might not.\label{ahb}}
\end{figure}

There are few ways to resolve the uncertainties regarding star formation
histories and the applicability of the \citet{calzetti00} attenuation formula
to dusty galaxies.  One (partial) way is to use the formula and
\alephha~to predict the escape fraction of the H$\beta$ luminosity,
\alephhb.  The ratio of intrinsic
H$\alpha$ to H$\beta$ luminosities is 2.86 under reasonable
assumptions; with H$\beta$ observations, we can use
($L_{\mathrm{H}\alpha}$/$L_{\mathrm{H}\beta}$)/2.86~=~\alephha/\alephhb~and from this
determine \alephhb.  If the predicted and determined H$\beta$ escape fractions
agree, the assumed attenuation formula is probably valid at optical
wavelengths; the assumed star formation histories are also probably valid.
The histories are valid because the prediction is history-dependent and the
determination is history-independent---both line luminosities are from ionized
gas surrounding the same stars.  If
the predicted and determined fractions disagree, the attenuation
formula at optical wavelengths can be different for dusty galaxies
or their true \alephha~can be different.  The formula can differ due to dust
not distributed as a uniform foreground screen; see table 1 and fig. 3b of
\citealt{calzetti01}.  Fig. \ref{ahb}
shows this test applied to the low-redshift dusty galaxies.  To apply this
test to high-redshift dusty galaxies, we need near-IR spectra that cover
the H$\beta$ line.  If the dispersion
of H$\alpha$ escape fractions at a given UV escape fraction for high-redshift
dusty galaxies remains, there must be a diversity of dust
distributions.  Either way, this test provides no information on the relation
between UV and 0.66$\,\micron$ escape fractions; more extrapolations are
necessary to definitively interpret Fig. \ref{fha_fuv}.

\section{Summary}\label{sec:conclude}

\begin{enumerate}
\item We measure H$\alpha$ line luminosities and limits, from
\emph{HST}/WFC3-IR grism spectra, for $z \sim 1$ galaxies with detected
100$\,\micron$ emission.  We determine rest-frame UV continuum power-law
indices ($\beta$ values) and ratios of IR to UV luminosities (IRX values).
\item For each galaxy we determine \alephuv, the escape fraction at
0.16$\,\micron$, from IRX.  We determine \alephha, the H$\alpha$ escape
fraction, from the H$\alpha$ luminosity and the total star formation rate.
\item For the \citet{overzier11} sample of low-redshift dust-poor galaxies, we
recover the \citet{calzetti97b} relation between H$\alpha$ and UV escape
fractions: \alephha~= \alephuv$^{0.80}$.  Our nonstandard method for determining
H$\alpha$ escape fractions should be grossly correct.
\item For the \citet{hwang13} sample of low-redshift dusty galaxies, we find
that \alephha~= \alephuv$^{0.55}$.
\item For our sample of dusty galaxies at $0.7 < z < 1.5$, we find that
\alephha~= \alephuv$^{0.55}$.  The dispersion about the best-fit relation is
large.
\item The interpretation of the results for dusty galaxies is unclear.  If we
assume that the \citet{calzetti00} attenuation formula applies to these
galaxies, our results agree with those of \citet{onodera10}, \citet{kashino13},
\citet{price13}, and \citet{reddy15} and conflict with those of \citet{erb06}: \alephha~=
\alephuvha$^{1.6}$.  They study
samples comprising dust-poor and dusty galaxies.  Dusty galaxies have
nonuniform dust distributions.  However, dusty galaxies may disobey the
\citet{calzetti00} attenuation formula.
\item Measurements of the ratios of H$\alpha$ to H$\beta$ luminosities may
decrease the dispersion in the relation between H$\alpha$ and UV escape
fractions for dusty galaxies at $z \sim 1$.  If the dispersion is real, these
galaxies have diverse dust distributions.
\end{enumerate}

ALMA is beginning to produce images of the dust distributions in high-redshift
galaxies.  These images lead to direct tests, independent of the attenuation
formula, of the uniform screen assumption.  They also have the potential to
reveal the parts of galaxies that may never be detected at UV and optical
wavelengths: surprises surely abound.

\clearpage

\begin{turnpage}
\begin{deluxetable}{llllllllllllll}
\tablecolumns{14}
\tablewidth{576.97723pt}
\tablecaption{Measured and derived quantities for the sample at $z > 0.7$
\label{tbl:highz}}
\tabletypesize{\tiny}
\tablehead{
\colhead{RA, IR} &
\colhead{Dec, IR} &
\colhead{RA, opt.} &
\colhead{Dec, opt.} &
\colhead{$z$} &
\colhead{$\beta$} &
\colhead{$\sigma_{\beta}$} &
\colhead{$0.16 \times L_{0.16}$} &
\colhead{$\sigma_{L_{0.16}}$} &
\colhead{$L_{\mathrm{H}\alpha}$} &
\colhead{$\sigma_{L_{\mathrm{H}\alpha}}$} &
\colhead{$L_{\mathrm{H}\alpha}$ lim.} &
\colhead{$L_{\mathrm{IR}}$} &
\colhead{$\sigma_{L_{\mathrm{IR}}}$} \\
\colhead{deg} &
\colhead{deg} &
\colhead{deg} &
\colhead{deg} &
\colhead{} &
\colhead{} &
\colhead{} &
\colhead{$L_{\odot}$} &
\colhead{$L_{\odot}$} &
\colhead{$L_{\odot}$} &
\colhead{$L_{\odot}$} &
\colhead{$L_{\odot}$} &
\colhead{$L_{\odot}$} &
\colhead{$L_{\odot}$}
}
\startdata
\cutinhead{GOODS-S}
53.113487 & -27.933195 & 53.113470 & -27.933294 & 1.098\tablenotemark{G} & 0.53 & 0.34 & 1.1e+10 & 4.9e+08 & 5.6e+08 & 2.1e+07 & \ldots & 6.2e+11 & 1.9e+10  \\
53.096371 & -27.925898 & 53.096385 & -27.925972 & 1.146 & -1.89 & 0.41 & 7.9e+09 & 3.3e+08 & 3.3e+08 & 2.6e+07 & \ldots & 2.2e+11 & 4.9e+10  \\
53.090958 & -27.922218 & 53.090977 & -27.922288 & 1.390 & 1.41 & 0.48 & 8.4e+08 & 4.5e+08 & 3.1e+08 & 2.2e+07 & \ldots & 5.5e+11 & 2.7e+10  \\
53.184750 & -27.920420 & 53.184801 & -27.920432 & 0.953 & -1.68 & 0.15 & 1.2e+10 & 3.9e+08 & 1.6e+08 & 2.3e+07 & \ldots & 9.3e+10 & 1.7e+07  \\
53.104836 & -27.913826 & 53.104830 & -27.913926 & 1.090 & 0.61 & 0.61 & 3.5e+09 & 2.7e+08 & 1.2e+08 & 1.7e+07 & \ldots & 1.8e+11 & 6.2e+09
\enddata
\tablenotetext{G}{Spectroscopic redshift solely from grism observations.}
\tablecomments{Columns: RA and Dec, IR, are positions in the
\emph{Herschel} catalogs.  They are based on positions of \emph{Spitzer}/IRAC
priors.  RA and Dec, opt., are positions in the CANDELS catalogs.
$L_{\mathrm{H}\alpha}$ lim. is a 5$\sigma$ upper limit to the H$\alpha$
line luminosity.  The entire table is published in the electronic edition of
\apj.  A portion is shown here for guidance regarding its form and content.}
\end{deluxetable}

\clearpage

\begin{deluxetable}{lllllllllllll}
\tablecolumns{13}
\tablewidth{493.65451pt}
\tablecaption{Measured and derived quantities for the samples at $z < 0.2$
\label{tbl:lowz}}
\tabletypesize{\tiny}
\tablehead{
\colhead{RA} &
\colhead{Dec} &
\colhead{$z$} &
\colhead{$\beta$} &
\colhead{$\sigma_{\beta}$} &
\colhead{$0.16 \times L_{0.16}$} &
\colhead{$\sigma_{L_{0.16}}$} &
\colhead{$L_{\mathrm{H}\beta}$} &
\colhead{$\sigma_{L_{\mathrm{H}\beta}}$} &
\colhead{$L_{\mathrm{H}\alpha}$} &
\colhead{$\sigma_{L_{\mathrm{H}\alpha}}$} &
\colhead{$L_{\mathrm{IR}}$} &
\colhead{$\sigma_{L_{\mathrm{IR}}}$} \\
\colhead{deg} &
\colhead{deg} &
\colhead{} &
\colhead{} &
\colhead{} &
\colhead{$L_{\odot}$} &
\colhead{$L_{\odot}$} &
\colhead{$L_{\odot}$} &
\colhead{$L_{\odot}$} &
\colhead{$L_{\odot}$} &
\colhead{$L_{\odot}$} &
\colhead{$L_{\odot}$} &
\colhead{$L_{\odot}$}
}
\startdata
\cutinhead{Dust-poor galaxies}
30.987125 & -8.132919 & 0.189 & -1.66 & 0.04 & 3.5e+10 & 2.9e+08 & 1.4e+08  & \ldots & 4.6e+08 & \ldots & 8.5e+10 & 1.8e+09 \\
33.452250 & 12.997628 & 0.219 & -0.45 & 0.05 & 4.3e+10 & 4.3e+08 & 3.6e+07  & \ldots & 1.4e+08 & \ldots & 5.4e+11 & 3.0e+09 \\
52.191625 & 1.197458 & 0.142 & -1.68 & 0.04 & 1.9e+10 & 1.7e+08 & 7.4e+07   & \ldots & 2.4e+08 & \ldots & 5.4e+10 & 1.0e+09 \\
59.391667 & -5.622139 & 0.204 & -1.28 & 0.06 & 2.9e+10 & 3.5e+08 & 9.4e+07  & \ldots & 3.5e+08 & \ldots & 1.1e+11 & 1.9e+09 \\
125.007167 & 50.844211 & 0.217 & -1.55 & 0.08 & 2.6e+10 & 4.0e+08 & 2.4e+08 & \ldots & 8.1e+08 & \ldots & 4.8e+11 & 4.9e+09
\enddata
\tablecomments{The entire table is published in the electronic edition of
\apj.  A portion is shown here for guidance regarding its form and content.}
\end{deluxetable}
\end{turnpage}
\clearpage


\begin{thebibliography}

\bibitem[Balestra et al.(2010)]{balestra10} Balestra, I., Mainieri, V., Popesso,
P., et al.\ 2010, \aap, 512, AA12

\bibitem[Boissier et al.(2004)]{boissier04} Boissier, S., Boselli, A.,
Buat, V., Donas, J., \& Milliard, B.\ 2004, \aap, 424, 465

\bibitem[Brammer et al.(2012)]{brammer12} Brammer, G.~B., van 
Dokkum, P.~G., Franx, M., et al.\ 2012, \apjs, 200, 13

\bibitem[Buat et al.(2012)]{buat12} Buat, V., Noll, S., Burgarella, D.,
et al.\ 2012, \aap, 545, A141 

\bibitem[Buat et al.(2011)]{buat11} Buat, V., Giovannoli, E., Heinis, S.,
et al.\ 2011, \aap, 533, A93 

\bibitem[Calzetti(2001)]{calzetti01} Calzetti, D.\ 2001, \pasp, 113, 1449

\bibitem[Calzetti et al.(2000)]{calzetti00} Calzetti, D., Armus, L., Bohlin,
R.~C., et al.\ 2000, \apj, 533, 682

\bibitem[Calzetti(1997b)]{calzetti97b} Calzetti, D.\ 1997b, American 
Institute of Physics Conference Series, 408, 403

\bibitem[Calzetti(1997a)]{calzetti97a} Calzetti, D.\ 1997a, \aj, 113, 162

\bibitem[Calzetti et al.(1994)]{calzetti94} Calzetti, D., Kinney, A.~L., \&
Storchi-Bergmann, T.\ 1994, \apj, 429, 582

\bibitem[Chary \& Elbaz(2001)]{chary01} Chary, R., \& Elbaz, D.\ 2001, \apj,
556, 562

\bibitem[D{\'{\i}}az-Santos et al.(2010)]{diaz10} D{\'{\i}}az-Santos, T.,
Charmandaris, V., Armus, L., et al.\ 2010, \apj, 723, 993

\bibitem[Donley et al.(2012)]{donley12} Donley, J.~L., Koekemoer, A.~M.,
Brusa, M., et al.\ 2012, \apj, 748, 142

\bibitem[Erb et al.(2006)]{erb06} Erb, D.~K., Steidel, C.~C., 
Shapley, A.~E., et al.\ 2006, \apj, 647, 128

\bibitem[Fadda et al.(2010)]{fadda10} Fadda, D., Yan, L., 
Lagache, G., et al.\ 2010, \apj, 719, 425

\bibitem[Galametz et al.(2013)]{galametz13} Galametz, A., Grazian, A.,
Fontana, A., et al.\ 2013, \apjs, 206, 10

\bibitem[Guo et al.(2013)]{guo13} Guo, Y., Ferguson, H.~C., 
Giavalisco, M., et al.\ 2013, \apjs, 207, 24

\bibitem[Howarth(1983)]{howarth83} Howarth, I.~D.\ 1983, \mnras, 203, 301

\bibitem[Hwang \& Geller(2013)]{hwang13} Hwang, H.~S., \& Geller, M.~J.\ 2013,
\apj, 769, 116

\bibitem[Juneau et al.(2013)]{juneau13} Juneau, S., Dickinson, 
M., Bournaud, F., et al.\ 2013, \apj, 764, 176

\bibitem[Kashino et al.(2013)]{kashino13} Kashino, D., Silverman, 
J.~D., Rodighiero, G., et al.\ 2013, \apjl, 777, L8

\bibitem[Kauffmann et al.(2003)]{kauffmann03} Kauffmann, G., 
Heckman, T.~M., Tremonti, C., et al.\ 2003, \mnras, 346, 1055

\bibitem[Kennicutt(1998)]{kennicutt98} Kennicutt, R., 1999, \araa, 36, 189

\bibitem[Kriek \& Conroy(2013)]{kriek13} Kriek, M., \& Conroy, C.\ 2013,
\apjl, 775, L16 

\bibitem[Kurk et al.(2013)]{kurk13} Kurk, J., Cimatti, A., Daddi, E., et al.\
2013, \aap, 549, AA63

\bibitem[Le F{\`e}vre et al.(2004)]{lefevre04} Le F{\`e}vre, O., Vettolani, G.,
Paltani, S., et al.\ 2004, \aap, 428, 1043

\bibitem[Lilly et al.(2009)]{lilly09} Lilly, S.~J., Le Brun, V.,
Maier, C., et al.\ 2009, \apjs, 184, 218

\bibitem[Liu et al.(2013)]{liu13} Liu, G., Calzetti, D., 
Hong, S., et al.\ 2013, \apjl, 778, L41

\bibitem[Madau \& Dickinson(2014)]{madau14} Madau, P., \& Dickinson, M.\ 2014,
arXiv:1403.0007

\bibitem[Magnelli et al.(2013)]{magnelli13} Magnelli, B., Popesso, P.,
Berta, S., et al.\ 2013, \aap, 553, A132

\bibitem[M{\'e}nard et al.(2010)]{menard10} M{\'e}nard, B., Scranton, R.,
Fukugita, M., \& Richards, G.\ 2010, \mnras, 405, 1025

\bibitem[Mignoli et al.(2005)]{mignoli05} Mignoli, M., Cimatti, A., Zamorani,
G., et al.\ 2005, \aap, 437, 883

\bibitem[Mu{\~n}oz-Mateos et al.(2009)]{munoz09} 
Mu{\~n}oz-Mateos, J.~C., Gil de Paz, A., Boissier, S., et al.\ 2009, \apj, 
701, 1965

\bibitem[Onodera et al.(2010)]{onodera10} Onodera, M., Arimoto, 
N., Daddi, E., et al.\ 2010, \apj, 715, 385

\bibitem[Overzier et al.(2011)]{overzier11} Overzier, R.~A., 
Heckman, T.~M., Wang, J., et al.\ 2011, \apjl, 726, L7 

\bibitem[Overzier et al.(2009)]{overzier09} Overzier, R.~A., 
Heckman, T.~M., Tremonti, C., et al.\ 2009, \apj, 706, 203

\bibitem[Penner et al.(2012)]{penner12} Penner, K., Dickinson, 
M., Pope, A., et al.\ 2012, \apj, 759, 28 

\bibitem[Prescott et al.(2007)]{prescott07} Prescott, M.~K.~M., 
Kennicutt, R.~C., Jr., Bendo, G.~J., et al.\ 2007, \apj, 668, 182

\bibitem[Price et al.(2013)]{price13} Price, S.~H., Kriek, M., 
Brammer, G.~B., et al.\ 2014, \apj, 788, 86

\bibitem[Ravikumar et al.(2007)]{ravikumar07} Ravikumar, C.~D., Puech, M.,
Flores, H., et al.\ 2007, \aap, 465, 1099

\bibitem[Reddy et al.(2015)]{reddy15} Reddy, N.~A., Kriek, M., 
Shapley, A.~E., et al.\ 2015, arXiv:1504.02782

\bibitem[Reddy et al.(2012)]{reddy12} Reddy, N., Dickinson, M.,
Elbaz, D., et al.\ 2012, \apj, 744, 154 

\bibitem[Reddy et al.(2010)]{reddy10} Reddy, N.~A., Erb, D.~K.,
Pettini, M., Steidel, C.~C., \& Shapley, A.~E.\ 2010, \apj, 712, 1070

\bibitem[Santini et al.(2014)]{santini14} Santini, P., Ferguson, 
H.~C., Fontana, A., et al.\ 2014, arXiv:1412.5180

\bibitem[Simpson et al.(2012)]{simpson12} Simpson, C., Rawlings, 
S., Ivison, R., et al.\ 2012, \mnras, 421, 3060

\bibitem[Simpson et al.(2015)]{simpson15} Simpson, J.~M., Smail, 
I., Swinbank, A.~M., et al.\ 2015, \apj, 799, 81

\bibitem[Szokoly et al.(2004)]{szokoly04} Szokoly, G.~P., 
Bergeron, J., Hasinger, G., et al.\ 2004, \apjs, 155, 271

\bibitem[Vanzella et al.(2008)]{vanzella08} Vanzella, E., Cristiani, S.,
Dickinson, M., et al.\ 2008, \aap, 478, 83

\bibitem[Weiner et al.(2007)]{weiner07} Weiner, B.~J., Papovich, 
C., Bundy, K., et al.\ 2007, \apjl, 660, L39

\bibitem[Willman et al.(2011)]{willman11} Willman, B., Geha, M., 
Strader, J., et al.\ 2011, \aj, 142, 128 

\bibitem[Willman et al.(2005)]{willman05} Willman, B., Blanton, 
M.~R., West, A.~A., et al.\ 2005, \aj, 129, 2692

\bibitem[Yamada et al.(2005)]{yamada05} Yamada, T., Kodama, T., 
Akiyama, M., et al.\ 2005, \apj, 634, 861

\bibitem[Younger et al.(2010)]{younger10} Younger, J.~D., Fazio, 
G.~G., Ashby, M.~L.~N., et al.\ 2010, \mnras, 407, 1268

\bibitem[Zaritsky(1994)]{zaritsky94} Zaritsky, D.\ 1994, \aj, 108, 
1619

\end{thebibliography}
\end{document}